# The Klein Paradox: A New Treatment

**E. Truebenbacher**

Institut fuer Physik, Johannes-Gutenberg-Universitaet, D-55128 Mainz, Germany

**E-mail address:**

trubegon@uni-mainz.de (E. Truebenbacher)

Abstract: The Dirac equation requires a treatment of the step potential that differs fundamentally from the traditional treatment, because the Dirac plane waves, besides momentum and spin, are characterized by a quantum number with the physical meaning of sign of charge. Since the Hermitean operator corresponding to this quantum number does not commute with the step potential, the time displacement parameter used in the ansatz of the stationary state does not have the physical meaning of energy. Therefore there are no paradoxal values of the "energy". The new solution of the Dirac equation with a step potential is obtained. This solution, again, allows for phenomena of the Klein paradox type, but in addition it contains a positron amplitude localized at the threshold point of the step potential.

**I) Introduction**

In this work the discussion of a particle moving in a step potential is resumed from a new point of view.

First, in Section II, it will be shown that the Dirac equation produces an observable "sign of charge" – briefly "charge operator" – which, together with momentum and spin, characterizes the plane wave solutions of the Dirac equation.

Section III discusses the introduction of a potential in the Dirac equation, including the two possible cases of potentials that do commute and potentials that do not commute with the charge operator. For the latter potentials simultaneous eigenfunctions of charge and energy do not exist. As a consequence the time displacement parameter of the stationary state used in the ansatz of the traditional solution, usually denoted by *E*, has no longer the physical meaning of energy. Since, in particular, the step potential does not commute with the charge operator this is true also for the stationary solution of the step problem. Therefore there are no paradoxal "energies".

In Section IV, the solution of the step potential is presented according to the theory developed in the preceding sections. This solution is clearly not an eigenfunction of the charge operator. The predictions of the new solution are compared with those of the traditional solution.

**II) The free Dirac equation and the operator *sign of charge***

Here we summarize the essentials of a one particle theory of the Dirac equation based on the existence of an Hermitean operator with the physical meaning of *sign of charge of the particle*, described in [1].

Let

$$D = -i\vec{\alpha}\cdot\vec{\partial} + \beta m = \begin{pmatrix} m & -i\vec{\sigma}\cdot\vec{\partial} \\ -i\vec{\sigma}\cdot\vec{\partial} & -m \end{pmatrix} \quad (1)$$

be the free Dirac Hamiltonian for an electron with mass $m$, and let $\Omega$ be the operator

$$\Omega = \sqrt{-\Delta + m^2} \ . \tag{2}$$

For simplicity we shall call the operator $\Omega$ the kinetic energy, although it contains the rest energy *m*.

$\Omega$ is a mathematically well-defined *tempered distribution* [2].

Because of

$$[D,\Omega] = 0 \tag{3}$$

the operator

$$sgn = \frac{D}{\Omega} = \left(-i\vec{\alpha}\cdot\vec{\partial} + \beta m\right)\Omega^{-1} = \Omega^{-1}\left(-i\vec{\alpha}\cdot\vec{\partial} + \beta m\right), \tag{4}$$

is well-known as the operator of *sign of frequency* [3], with eigenvalues $q = \pm 1$.

Because of (3) we also have

$$[sgn,\Omega] = 0 \tag{5a}$$

$$[sgn,D] = 0 \tag{5b}$$

From the property

$$sgn\,\hat{C} = -\hat{C}\,sgn, \tag{6}$$

where $\hat{C}$ means the *charge conjugation transformation*, defined by

$$\hat{C}\Psi = \Psi' = C\Psi^*, \tag{7}$$

*C* being the matrix

$$C = \begin{pmatrix} 0 & -i\sigma_y \\ i\sigma_y & 0 \end{pmatrix} \tag{8}$$

it follows that the operator *sgn* has *the physical meaning of* "*sign of charge of the particle*".

To show this, suppose $\Psi$ is an eigenfunction of the operator (4), with eigenvalue *q*,

$$sgn\ \Psi = q\,\Psi, \quad q = \pm 1\ ,$$

With the help of (6) we find

$$sgn\,(\hat{C}\Psi) = -\hat{C}(sgn\,\Psi) = -\hat{C}(q\Psi) = -q(\hat{C}\Psi)\ ,$$

This means:

$\hat{C}$ *reverses the sign of the eigenvalue of the operator sgn*.

Since we know that $\hat{C}$ reverses the *sign of the charge* of the particle we *must identify the eigenvalues of the operator sgn with the sign of charge*,

$$sgn = sign\ of\ charge\ , \qquad q.e.d.$$

We shall call the operator *sgn* simply the *charge operator*.

As we can see from (4) the free *Dirac equation*

$$i\partial_t \Psi = D\Psi = sgn \cdot \Omega \Psi \qquad (9)$$

is just *the zero component* of the 4-*component dynamical relation*

$$\boxed{i\partial^\mu = sgn \cdot P^\mu} \qquad (10)$$

where $P^\mu = \{P^0, \vec{P}\}$ is *the four-momentum*, with the zero component $P^0 = \Omega$.

Therefore due to the Dirac equation *and relativistic covariance* we must *replace* the familiar non-relativistic relations

$$P^\mu = i\partial^\mu, \qquad (11)$$

by the relations (10). The relations (10) *must be used with Dirac wave functions*, instead of the nonrelativistic relations (11).

Solving (10) for $P^\mu$ by multiplying both sides by *sgn* and using

$$sgn^2 = 1$$

we have

$$\boxed{P^\mu = sgn \cdot i\partial^\mu}. \qquad (12)$$

The relations (12) express the statement of R. P. Feynman [4]: *"Antiparticles run backward in time."*

Clearly

$$[sgn, P^\mu] = 0 \qquad (13)$$

The well-known plane wave solutions of the Dirac equation for positive and negative frequencies, resp.

$$\Psi_+^{(s)}(t,\vec{x};\vec{k}) = \sqrt{\frac{\omega+m}{2\omega}} \begin{pmatrix} \chi^s \\ \frac{\vec{k}\cdot\vec{\sigma}}{\omega+m}\chi^s \end{pmatrix} \frac{1}{\sqrt{2\pi}^3} \exp(-i\omega t + i\vec{k}\cdot\vec{x}) \qquad (14a)$$

$$\Psi_-^{(s)}(t,\vec{x};-\vec{k}) = \sqrt{\frac{\omega+m}{2\omega}} \begin{pmatrix} \frac{\vec{k}\cdot\vec{\sigma}}{\omega+m}\chi^s \\ \chi^s \end{pmatrix} \frac{1}{\sqrt{2\pi}^3} \exp(+i\omega t - i\vec{k}\cdot\vec{x}), \qquad (14b)$$

$$s = \pm 1: \quad \chi^{+1} = \begin{pmatrix} 1 \\ 0 \end{pmatrix}, \; \chi^{-1} = \begin{pmatrix} 0 \\ 1 \end{pmatrix},$$

are *simultaneous eigenfunctions* of the three observables *charge sgn*, *momentum* $\vec{P}$ and *rest spin*, with quantum numbers $q = \pm 1$, $\vec{k}$ and $s$ resp.:

$$sgn\, \Psi_\pm^{(s)} = \pm \Psi_\pm^{(s)},$$

$\vec{P} = sgn \cdot (-i\vec{\partial})$ from (11) gives

$$\vec{P}\Psi_\pm^{(s)} = sgn\cdot(-i\vec{\partial})\Psi_\pm^{(s)} = \vec{k}\,\Psi_\pm^{(s)},$$

and for the quantum number $s = \pm 1$ we use the word *rest spin* of the particle since it is the value of $\sigma_z$ in the rest frame $\vec{k} = 0$. It is easy to construct the corresponding operator, but we shall not need it.

With respect to the *energy* $P^0 = \Omega$ we find, using

$$\Omega e^{-i\omega t + i\vec{k}\cdot\vec{x}} = \omega(\vec{k})\, e^{-i\omega t + i\vec{k}\cdot\vec{x}} \qquad (15)$$

the *positive value* $\omega(\pm\vec{k}) = \sqrt{\vec{k}^2 + m^2}$ in either case $\Psi_\pm^{(s)}$:

$$P^0 \Psi_\pm^{(s)}(t,\vec{x};\vec{k}) = \Omega \Psi_\pm^{(s)}(t,\vec{x};\vec{k}) = \omega(\vec{k})\Psi_\pm^{(s)}(t,\vec{x};\vec{k}), \qquad (16)$$

in agreement with (11),

$$P^0 \Psi_\pm^{(s)}(t,\vec{x};\vec{k}) = sgn\cdot i\partial_t \Psi_\pm^{(s)}(t,\vec{x};\vec{k}).$$

The *sign of the frequency* is *the sign of the charge*, not "the sign of the energy".

Thus the charge operator reveals the Dirac equation as *a one-particle equation for a particle with two charge states*. The operator *sgn* appears as the natural complement of the operator $\hat{C}$ and makes the the theory of the Dirac equation a consistent theory, perfectly symmetric in electrons and positrons, without the necessity of introducing a Dirac sea and hole states.

An arbitrary Dirac wave function $\Psi(t,\vec{x})$ may be expanded in the complete set of functions

$$\Psi^{(+,s)}(\vec{x};\vec{k}) = N(\vec{k}) \begin{pmatrix} \chi^s \\ \dfrac{\vec{k}\cdot\vec{\sigma}}{\omega+m}\chi^s \end{pmatrix} e^{i\vec{k}\cdot\vec{x}} = \tilde{\Psi}^{(+,s)}(\vec{k})e^{i\vec{k}\cdot\vec{x}}, \qquad (17a)$$

$$\Psi^{(-,s)}(\vec{x};\vec{k}) = N(\vec{k}) \begin{pmatrix} \dfrac{\vec{k}\cdot\vec{\sigma}}{\omega+m}\chi^s \\ \chi^s \end{pmatrix} e^{-i\vec{k}\cdot\vec{x}} = \tilde{\Psi}^{(-,s)}(\vec{k})e^{-i\vec{k}\cdot\vec{x}} \qquad (17b)$$

with evident meaning of $\tilde{\psi}^{(\pm,s)}$, with time dependent coefficients $a^{(\sigma,s)}(t;\vec{k})$:

$$\Psi(t,\vec{x}) = \sum_{\sigma s}\int dk^3\, a^{(\sigma,s)}(t;\vec{k})\, \tilde{\Psi}^{(\sigma,s)}(\vec{k}) e^{\sigma(i\vec{k}\cdot\vec{x})} = \Psi^{(+)}(t,\vec{x}) + \Psi^{(-)}(t,\vec{x}) =$$

$$= \Psi^{(ele)}(t,\vec{x}) + \Psi^{(pos)}(t,\vec{x}), \qquad (18)$$

where

$$\Psi^{(ele)}(t,\vec{x}) = \Psi_+(t,\vec{x}) = \sum_s \int dk^3\, a^{(+,s)}(t;\vec{k})\, \tilde{\Psi}^{(+,s)}(\vec{k}) e^{+i\vec{k}\cdot\vec{x}} \qquad (19a)$$

$$\Psi^{(pos)}(t,\vec{x}) = \Psi_-(t,\vec{x}) = \sum_s \int dk^3\, a^{(-,s)}(t;\vec{k})\, \tilde{\Psi}^{(-,s)}(\vec{k}) e^{-i\vec{k}\cdot\vec{x}} \qquad (19b)$$

are pure electron and positron functions resp., (see also [5], page 549 at the bottom).

Remark.

Dirac in his textbook on quantum mechanics says [6]: "We cannot simply assert that the negative energy solutions represent positrons, as this would make the dynamical relations wrong". We have seen that, indeed, we must retain the negative energy solutions as positrons, but it is the dynamical relations (11) that we must generalize by (12).

This seems the suitable place to make a remark concerning the statement in the textbook of Bjorken-Drell [7] that '*In the free field case . . . for the positron function we just obtain another electron function*': The author believes that the definition of the positron function should not give up orthogonality of particles with opposite charges.

The construction of a positron wave function in consistency with the rules of multifermion wave functions would require the construction of the Slater determinant for a hole state. Since this is impossible one depends on definitions. Dirac rules out the definition of positrons as negative energy solutions because of violation of the dynamical relations. The authors of [7] make use of the scope left for definitions to define the positron functions as the charge conjugates of the negative energy electron functions, thus accepting violation of orthogonality of particles with opposite charges. This case is not taken in consideration by Dirac.

**III) Dirac equation with a potential**

Adding a potential $V(\vec{x})$ to the free Dirac equation requires great care, because the Dirac wave function bears the quantum number *sign of charge*.

Multiplying (9) by $sgn$ we write the free Dirac equation as

$$i\partial_t sgn \Psi = \Omega \Psi \tag{20}$$

Remember that according to (18) a general Dirac wave function consists of two components with quantum numbers $sgn = +1$ and $sgn = -1$ resp.,

$$\Psi(t,\vec{x}) = \Psi^{(ele)}(t,\vec{x}) + \Psi^{(pos)}(t,\vec{x}),$$

Taking into account that the potential energy of the particle depends on the sign of its charge, i.e.

$$\left(E_{pot}\right)_{ele} = V(\vec{x}), \tag{21a}$$

$$\left(E_{pot}\right)_{pos} = -V(\vec{x}), \tag{21b}$$

we put

$$E_{pot} = V(\vec{x}) \cdot sgn \tag{22}$$

A potential energy should be added to the kinetic energy $\Omega$, therefore in (20) we replace $\Omega$ by $\Omega + V(\vec{x}) \cdot sgn$,

$$i\partial_t sgn \Psi = \{\Omega + V(\vec{x}) \cdot sgn\}\Psi = P^0 \Psi, \qquad (23)$$

$P^0$ being the total energy,

$$P^0 = \Omega + V \cdot sgn \qquad (24)$$

To remove $sgn$ from the left hand side in (23) we multiply from the left by $sgn$ and obtain

$$\boxed{i\partial_t \Psi = (D + sgn \cdot V(\vec{x}) \cdot sgn)\Psi} \qquad (25)$$

Since the potential is time *independent*, *i.e.*

$$[i\partial_t, sgn \cdot V(\vec{x}) \cdot sgn] = 0, \qquad (26)$$

there exist *stationary states*,

$$\Psi(t,\vec{x}) = e^{-i\eta t}\Phi(\vec{x}) \qquad (27)$$

Substituting (27) in (25) we obtain for the space part $\Phi(\vec{x})$

$$(D + sgn \cdot V(\vec{x}) \cdot sgn)\Phi = \eta \Phi \qquad (28a)$$

briefly

$$(D + V')\Phi = \eta \Phi \qquad (28b)$$

where

$$V' = sgn \cdot V \cdot sgn$$

We must distinguish the two cases of potentials that commute with $sgn$, and potentials that doe not commute with $sgn$.

Case1)

Let

$$[sgn, V(\vec{x})] = 0, \qquad (29)$$

and therefore

$$V'(\vec{x}) = V(\vec{x}) \qquad (30)$$

Equation (28b) reduces to

$$(D+V)\Phi = \eta \Phi \qquad (31)$$

Since because of (29) and (5b) we also have

$$[sgn, D+V] = 0$$

there exist *simultaneous eigenfunctions* $\Phi$ of $sgn$ and $D+V$,

$$sgn\, \Phi = \sigma \Phi, \qquad \sigma = \pm 1 \qquad (32)$$

$$(D+V)\Phi = \eta \Phi$$

Using

$$D+V = P^0 sgn \qquad (33)$$

we finally arrive at

$$\boxed{P^0 \Phi(\vec{x}) = \sigma \eta\, \Phi(\vec{x})} \qquad (34)$$

(34) says that $\Phi(\vec{x})$ is an *eigenfunction of the energy* $P^0$ with the *eigenvalue* $\sigma\eta$. For electrons $\sigma = 1$, therefore the time displacement parameter $\eta$ has *the physical meaning of energy.*

Let us call wave functions satisfying (32) *pure charge states*, and potentials satisfying (29) *charge conserving potentials*.

Case 2)

Suppose

$$[sgn, V(\vec{x})] \neq 0 \qquad (35a)$$

and therefore

$$[sgn, V'(\vec{x})] \neq 0 \qquad (35b)$$

Now the solution of (28b) *cannot be an eigenfunction of* $sgn$, *i.e.* a pure charge state. Since the (one-dimensional) step potential at hand is an example of a potential that obviously does *not* commute with the operator $sgn$,

$$[V_0 \cdot \Theta(z), sgn] = \left[V_0 \cdot \Theta(z), \frac{D}{\Omega}\right] = V_0 \left[\Theta(z), \frac{-i\alpha_z \partial_z + \beta m}{\Omega}\right] \neq 0, \qquad (36)$$

let us generally discuss the solutions for non-commuting potentials.

From (28b) we now obtain, instead of (31),

$$(D+V)(sgn\Phi) = \eta(sgn\Phi) \tag{37}$$

and the identity (33) gives

$$P^0\Phi = \eta(sgn\Phi)$$

and therefrom

$$(sgn\ P^0)\Phi(\vec{x}) = \eta\Phi(\vec{x}) \tag{38}$$

Equation (37) is identical with the stationary Dirac equation of the traditional solution $\Phi_{trad}(z)$, say, with $sgn\Phi$ in place of $\Phi_{trad}(z)$ and $\eta$ in place of $E$.

Equation (38) means that the *eigenvalue $\eta$ of the time displacement operator $i\partial_t$* now is the eigenvalue of the operator $sgn\ P^0$, but not of the energy $P^0$. Therefore for potentials that do not commute with charge $\eta$ *does not have the physical meaning of the energy*.

Remark:

The time displacement parameter $\eta$ is related to the *expectation value* of the energy:

Let

$$\Psi(t,\vec{x}) = e^{-i\eta t}\Phi(\vec{x})$$

Using (23), we obtain for the expectation value of $P^0$ in the state $\Psi(t,\vec{x})$:

$$\langle\Psi(t,\vec{x})|P^0|\Psi(t,\vec{x})\rangle = \langle\Psi(t,\vec{x})|sgn\cdot i\partial_t|\Psi(t,\vec{x})\rangle = \eta\langle\Phi(\vec{x})|sgn|\Phi(\vec{x})\rangle = \eta\langle sgn\rangle$$

and therefore the relation

$$\langle P^0\rangle = \langle sgn\rangle \cdot \eta \tag{39}$$

If the potential energy $E_{pot} = V(\vec{x})\cdot sgn$ does not commute with the operator $sgn$ the wave function consists of two components with quantum numbers $sgn = +1$ and $sgn = -1$ resp., see (18),

$$\Psi(t,\vec{x}) = \Psi^{(ele)}(t,\vec{x}) + \Psi^{(pos)}(t,\vec{x})$$

A method to construct potentials that exactly commute with $sgn$ is described in Section 7.2 of reference [1].

Up to this point the considerations are general.

## IV) The step potential

We know that the step potential does not commute with the charge operator $sgn$, (36), and therefore we are faced with the case 2) of the preceding section. As a consequence the traditional solution $\Phi_{trad}(z)$ is *not as yet the final solution* $\Phi_{step}(z)$, say, of the step problem.

As the simplest solution let us identify $sgn\Phi_{step}$ with $\Phi_{trad}(z)$,

$$\Phi_{trad}(z) = sgn\,\Phi_{step}(z) \tag{40a}$$

Due to (40a) the final solution $\Phi_{step}(z)$ is obtained by *applying the operator $sgn$ on $\Phi_{trad}(z)$*

$$\Phi_{step}(z) = sgn\,\Phi_{trad}(z) \tag{40b}$$

We compile for the traditional solution

$$\psi_{trad} = \left(a\tilde{\Psi}_p e^{ipz} + c\tilde{\Psi}_{-p} e^{-ipz}\right)\cdot(1-\Theta(z)) + b\tilde{\Psi}_q e^{iqz}\cdot\Theta(z) \tag{41}$$

$$V(z) = V_0 \cdot \Theta(z) \tag{42}$$

$$\Theta(z) = \begin{cases} 0 & for\ z \le 0 \\ 1 & for\ z > 0 \end{cases} \tag{43}$$

$$\Theta(z)(1-\Theta(z)) = \begin{cases} 0 & for\ x \le 0 \\ 0 & for\ x > 0 \end{cases} = 0 \tag{44a}$$

$$\Theta(z)\cdot\Theta(z) = \Theta(z) \tag{44b}$$

$$\Theta'(z) = \delta(z) \tag{45a}$$

$$\Theta''(z) = \delta'(z) \tag{45b}$$

$$f(z)\delta(z) = f(0)\delta(z) \tag{46a}$$

$$f(z)\delta'(z) = -f'(0)\delta(z) \tag{46b}$$

$$\tilde{\Psi}_p = \frac{1}{\sqrt{2\pi}}\sqrt{\frac{\omega_p + m}{2\omega_p}}\begin{pmatrix} 1 \\ 0 \\ \frac{p\sigma_z}{\omega_p + m}\begin{pmatrix} 1 \\ 0 \end{pmatrix} \end{pmatrix} = \frac{1}{\sqrt{2\pi}}\sqrt{\frac{\omega_p + m}{2\omega_p}}\begin{pmatrix} 1 \\ 0 \\ \frac{p}{\omega_p + m} \\ 0 \end{pmatrix} = N_p \begin{pmatrix} \chi \\ \frac{p}{\omega_p + m}\chi \end{pmatrix} \tag{47}$$

$$\chi = \begin{pmatrix} 1 \\ 0 \end{pmatrix}$$

$$\omega_p = \sqrt{p^2 + m^2}$$

$$N_p = \frac{1}{\sqrt{2\pi}}\sqrt{\frac{\omega_p + m}{2\omega_p}}$$

Correspondingly, for momenta $-p, q$ .                ,

We must evaluate, according to (40b),

$$\text{sgn}\,\Phi_{trad}(z) = \left(-i\vec{\alpha}\cdot\vec{\partial} + \beta m\right)\frac{1}{\Omega}\left\{\left(a\tilde{\Psi}_p e^{ipz} + c\tilde{\Psi}_{-p}e^{-ipz}\right)\cdot\left(1 - \Theta(z)\right) + b\tilde{\Psi}_q e^{iqz}\cdot\Theta(z)\right\}$$

We calculate for a general function $f(\Omega)$:

$$f(\Omega)\left(e^{ipz}\cdot(1-\Theta(z))\right) = \int_{-\infty}^{+\infty} dz'\left(\frac{1}{2\pi}\int_{-\infty}^{+\infty} dk\, f(\omega(k))e^{ik\cdot(z-z')}\right)\left(e^{ipz'}\cdot(1-\Theta(z'))\right) =$$

$$= \frac{1}{2\pi}\int_{-\infty}^{+\infty} dk\, f(\omega(k))e^{ikz}\left[\int_{-\infty}^{+\infty} dz'\, e^{-i(k-p)z'}\cdot(1-\Theta(z'))\right] =$$

$$= \frac{1}{2\pi}\int_{-\infty}^{+\infty} dk\, f(\omega(k))e^{ikz}\left[\int_{-\infty}^{0} dz'\, e^{-i(k-p)z'}\right] =$$

$$= \frac{1}{2\pi}\int_{-\infty}^{+\infty} dk\, f(\omega(k))e^{ikz}\cdot 2\pi\delta_+(k-p), \tag{48}$$

where the distribution $\delta_+(k)$ is defined by [8].

$$\delta_+(k) = \frac{1}{2\pi}\int_{-\infty}^{0} dz\, e^{-ikz}, \tag{49}$$

According to [8] $\delta_+(k)$ may be expressed as

$$\delta_+(k) = -\frac{1}{2\pi i}\frac{1}{k+i\varepsilon} \tag{50}$$

which gives

$$f(\Omega)\left(e^{ipz}\cdot(1-\Theta(z))\right) = \int_{-\infty}^{+\infty} dk\ f(\omega(k))e^{ikz}\left(-\frac{1}{2\pi i}\frac{1}{k-(p-i\varepsilon)}\right)$$

In our case $f(\omega(k))$ has no pole in the complex $k$-plane, therefore the only pole of the integrand is in $k = p - i\varepsilon$.

If $z > 0$, we can complete the path in the $k$-plane from $-\infty$ to $+\infty$ by a semicircle in the upper half plane. There is no pole, therefore, according to the residue theorem we obtain

$$f(\Omega)\left(e^{ipz}\cdot(1-\Theta(z))\right) = \int_{-\infty}^{+\infty} dk\ f(\omega(k))e^{ikz}\left(-\frac{1}{2\pi i}\frac{1}{k-(p-i\varepsilon)}\right) = 0 \quad \text{for } z > 0$$

If $z < 0$, we can complete the path in the $k$-plane from $-\infty$ to $+\infty$ by a semicircle in the lower half plane. There is a pole in $k = (p-i\varepsilon)$. Since the direction of rotation is now negative, we have

$$f(\Omega)\left(e^{ipz}\cdot(1-\Theta(z))\right) = \int_{-\infty}^{+\infty} dk\ f(\omega(k))e^{ikz}\left(-\frac{1}{2\pi i}\frac{1}{k-(p-i\varepsilon)}\right) = f(\omega(p))e^{ipz} \quad \text{for } z < 0$$

For $z = 0$ we cannot complete the path by semicircles because the factor $e^{ikz}$ creating zero for the contribution of the semicircles is now absent. Thus we are left with

$$f(\Omega)\left(e^{ipz}\cdot(1-\Theta(z))\right)\Big|_{z=0} = \int_{-\infty}^{+\infty} dk\ f(\omega(k))\left(-\frac{1}{2\pi i}\frac{1}{k-(p-i\varepsilon)}\right)$$

Let us abbreviate the formal expression on the right hand side by

$$-\frac{1}{2\pi i}\int_{-\infty}^{+\infty} dk\ \frac{1}{k-(p-i\varepsilon)}f(\omega(k)) = f_+(p) \tag{51}$$

We formally combine the preceding results in the formula

$$f(\Omega)\left(e^{ipz}\cdot(1-\Theta(z))\right) = f(\omega(p))e^{ipz}\cdot(1-\Theta(z)) + \left[-f(\omega(p)) + f_+(p)\right]\cdot\frac{\delta(z)}{\delta(0)} \tag{52}$$

Replacing $p$ by $-p$ gives

$$f(\Omega)\left(e^{-ipz} \cdot (1 - \Theta(z))\right) = f(\omega(p))e^{-ipz} \cdot (1 - \Theta(z)) + \left[-f(\omega(p)) + f_+(-p)\right] \cdot \frac{\delta(z)}{\delta(0)} \qquad (53)$$

Similarly we find

$$f(\Omega)\left(e^{iqz} \cdot \Theta(z)\right) = \frac{1}{2\pi} \int_{-\infty}^{+\infty} dk\, f(\omega(k))e^{ikz} \cdot 2\pi \delta_-(k-q), \qquad (54)$$

where the distribution $\delta_-(k)$ is defined by [8]

$$\delta_-(k) = \frac{1}{2\pi} \int_0^\infty dz\, e^{-ikz}, \qquad (55)$$

According to [8] $\delta_-(k)$ may be expressed as

$$\delta_-(k) = \frac{1}{2\pi i} \frac{1}{k - i\varepsilon} \qquad (56)$$

which gives

$$f(\Omega)\left(e^{iqz} \cdot \Theta(z)\right) = \int_{-\infty}^{+\infty} dk\, f(\omega(k))e^{ikz} \left(\frac{1}{2\pi i} \frac{1}{k - (q + i\varepsilon)}\right) \qquad (57)$$

$f(\omega(k))$ has no pole in the complex $k$-plane, therefore the only pole of the integrand is in $k = q + i\varepsilon$.

For $z > 0$, we complete the path in the $k$-plane from $-\infty$ to $+\infty$ in the upper half plane, for $z < 0$, we complete the path in the $k$-plane from $-\infty$ to $+\infty$ in the lower half plane.

For $z = 0$ we are left with

$$f(\Omega)\left(e^{ipz} \cdot \Theta(z)\right)\Big|_{z=0} = \int_{-\infty}^{+\infty} dk\, f(\omega(k)) \left(\frac{1}{2\pi i} \frac{1}{k - (q + i\varepsilon)}\right)$$

Abbreviating the formal expression on the right hand side by

$$\frac{1}{2\pi i} \int_{-\infty}^{+\infty} dk\, \frac{1}{k - (q + i\varepsilon)} f(\omega(k)) = f_-(q) \qquad (58)$$

we combine the results in the formula

$$f(\Omega)\left(e^{iqz}\cdot\Theta(z)\right)=f(\omega(q))e^{iqz}\cdot\Theta(z)+f_{-}(q)\frac{\delta(z)}{\delta(0)} \qquad (59)$$

We summarize

$$f(\Omega)\left(e^{ipz}\cdot(1-\Theta(z))\right)=f(\omega(p))e^{ipz}\cdot(1-\Theta(z))+\left[-f(\omega(p))+f_{+}(p)\right]\cdot\frac{\delta(z)}{\delta(0)} \qquad (60a)$$

$$f(\Omega)\left(e^{-ipz}\cdot(1-\Theta(z))\right)=f(\omega(p))e^{-ipz}\cdot(1-\Theta(z))+\left[-f(\omega(p))+f_{+}(-p)\right]\cdot\frac{\delta(z)}{\delta(0)} \qquad (60b)$$

$$f(\Omega)\left(e^{iqz}\cdot\Theta(z)\right)=f(\omega(q))e^{iqz}\cdot\Theta(z)+f_{-}(q)\frac{\delta(z)}{\delta(0)} \qquad (60c)$$

In our case $f(\Omega)=\frac{1}{\Omega}$.

With the help of (60a) we find for $\Phi_{trad}(z)$ from (41)

$$\begin{aligned}\frac{1}{\Omega}\Phi_{trad}(z)=&\ a\tilde{\psi}_{p}\left[\frac{1}{\omega_{p}}e^{ipz}(1-\Theta(z))+\left(-\frac{1}{\omega_{p}}+f_{+}(p)\right)\frac{\delta(z)}{\delta(0)}\right]+\\ &+c\tilde{\psi}_{-p}\left[\frac{1}{\omega_{p}}e^{-ipz}(1-\Theta(z))+\left(-\frac{1}{\omega_{p}}+f_{+}(-p)\right)\frac{\delta(z)}{\delta(0)}\right]+\\ &+b\tilde{\psi}_{q}\left[\frac{1}{\omega_{q}}e^{iqz}\Theta(z)+f_{-}(p)\frac{\delta(z)}{\delta(0)}\right]\end{aligned} \qquad (61)$$

On expression (61) we must apply $D=-i\alpha_{z}\partial_{z}+\beta m$.

The $a$-term:

$$(-i\alpha_{z}\partial_{z}+\beta m)\left\{\tilde{\Psi}_{p}\left[\frac{1}{\omega_{p}}e^{ipz}(1-\Theta(z))+\left(-\frac{1}{\omega_{p}}+f_{+}(p)\right)\frac{\delta(z)}{\delta(0)}\right]\right\}=$$

$$=(\alpha_{z}p+\beta m)\frac{1}{\omega_{p}}\tilde{\Psi}_{p}e^{ipz}(1-\Theta(z))+\alpha_{z}\tilde{\Psi}_{p}\frac{1}{\omega_{p}}e^{ipz}i\delta(z)+$$

$$+\alpha_{z}\tilde{\Psi}_{p}\left(-\frac{1}{\omega_{p}}+f_{+}(p)\right)\frac{1}{\delta(0)}(-i\delta'(z))+\beta m\tilde{\Psi}_{p}\left(-\frac{1}{\omega_{p}}+f_{+}(p)\right)\frac{1}{\delta(0)}\delta(z)$$

Using

$$(\alpha_z p + \beta m)\frac{1}{\omega_p}\tilde{\Psi}_p = \tilde{\Psi}_p$$

and, because of $\sigma_z \chi = \chi$,

$$\alpha_z \tilde{\Psi}_p = \begin{pmatrix} & \sigma_z \\ \sigma_z & \end{pmatrix} N_p \begin{pmatrix} \chi \\ \frac{p}{\omega_p + m}\chi \end{pmatrix} = N_p \begin{pmatrix} \frac{p}{\omega_p + m}\chi \\ \chi \end{pmatrix} = \tilde{\Psi}_p^{po},$$

where $\tilde{\Psi}_p^{po}$ denotes the positron plane wave with momentum $p$, and $f(z)\delta(z) = f(0)\delta(z)$, we obtain for the *a*-contribution

$$\tilde{\Psi}_p e^{ipz}(1-\Theta(z)) +$$

$$+ \left\{\beta m \frac{1}{\delta(0)}\left(-\frac{1}{\omega_p} + f_+(p)\right)\tilde{\Psi}_p + i\frac{1}{\omega_p}\tilde{\Psi}_p^{po}\right\}\delta(z) - i\frac{1}{\delta(0)}\left(-\frac{1}{\omega_p} + f_+(p)\right)\tilde{\Psi}_p^{po}\delta'(z) \quad (62a)$$

The *c*-contribution is obtained by replacing $p$ by $-p$.

For the *b*-contribution we find

$$\tilde{\Psi}_q e^{iqz}\Theta(z) +$$

$$+ \left\{\beta m \tilde{\Psi}_q \frac{1}{\delta(0)}f_-(q) - i\frac{1}{\omega_p}\tilde{\Psi}_p^{po}\right\}\delta(z) - i\tilde{\Psi}_q^{po}f_-(q)\frac{1}{\delta(0)}\delta'(z) \quad (62b)$$

Adding the three contributions we obtain for the final solution

$$\boxed{\Phi_{step}(z) = \Phi_{trad}(z) + \ldots,} \quad (63)$$

where the dots denote the sum of the remaining terms. These terms are all proportional to $\delta$-functions and its derivative at the threshold $z = 0$. It is these terms by which $\Phi_{step}(z)$ differs from $\Phi_{trad}(z)$ and which contain the positron admixture. Without the $\delta$-function terms $\Phi_{trad}$ would be an eigenfunction of *sgn* with eigenvalue $+1$.

As far as the parameter $\eta$ is concerned it is hardly possible to find a physical meaning beyond (39). As a consequence, there is little use of talking about paradoxal values of the "energy".

All these results follow straight forward from the Dirac equation and requirements of relativistic covariance only, without additional assumptions or extensions.

Let us conclude with the following remark:

Y. V. Kononets [5] introduces the operator "*charge-index observable*". In the case of absence of potentials this observable is identical with the present operator of *sign of charge*. This author says that the correct treatment of the step potential "forbids the physical phenomena of the Klein paradox type", (see "Abstract" and "Concluding Remarks"). However, Y. V. K.'s method amounts to inserting the step potential in the Foldy-Wouthuysen picture of the Dirac equation. There it commutes with the operator *sign of charge* and therefore does so after transformation back into the Dirac picture. But the resulting Dirac picture potential is no longer the step potential, but a different (non-local) potential that does commute with the charge operator and therefore leads to solutions without Klein phenomena. (See in this context Section 7.2 of reference .)

Recently there obviously was observed experimental support of the Klein paradox [9]. Therefore it is important to observe: The solution (63) differs from the traditional solution only by the delta-function contributions. The traditional term, as before, does *not* forbid Klein phenomena. The delta-function part contains a non-vanishing positron amplitude. This would mean the appearance of positrons localized at the threshold point $z = 0$ of the step potential.

**List of References**